\documentclass[aps,prl,showpacs,twocolumn,groupedaddress]{revtex4-2}
\usepackage{amsmath}
\usepackage{amsfonts}
\usepackage{amssymb}
\usepackage{graphicx}
\usepackage{verbatim}
\usepackage{xcolor}
\usepackage{bbold}
\usepackage[normalem]{ulem}
\newcommand{\ts}{\textsuperscript}

\begin{document}

\title{Mechanisms of high harmonic generation in solids}

\author{A. Thorpe\ts{1}}
\email[]{athor087@uottawa.ca}
\author{N. Boroumand\ts{1}}
\email[]{nboro046@uottawa.ca}
\author{A. M. Parks\ts{1}}
\author{E. Goulielmakis\ts{2}}
\author{T. Brabec\ts{1}}

\affiliation{\ts{1}Department of Physics, University of Ottawa, Ottawa, ON K1N 6N5, Canada}
\affiliation{\ts{2}Institute of Physics, University of Rostock, Rostock, D-18059, Germany}

\date{\today}

\begin{abstract}
\noindent

The long standing issue of separating resonant from non-resonant processes in extreme nonlinear optics is resolved. The theoretical formalism is applied to high harmonic generation 
(HHG) in solids and reveals a deeper view into the dominant laser and material dependent mechanisms. Mid-infrared driven HHG in semiconductors is dominated by the resonant interband 
current. As a result of the dynamic Stark shift, virtual processes gain in importance in near-infrared driven HHG in dielectrics. Finally, our analysis identifies limitations of 
microscopic one-electron-hole theories. 
\end{abstract}

\maketitle

\noindent
Since its first demonstration a decade ago \cite{ghimire11}, high harmonic generation (HHG) in solids has opened novel avenues to probe material properties, such as the crystal momentum 
dependent bandgap \cite{vampa15r,luu15,garg16,garg18,lanin17}, Berry phase \cite{liu17,luu18}, and valence potential \cite{lakhotia20}; further, the sensitivity of HHG to lattice asymmetry 
\cite{jiang18}, to topological properties and correlation of materials \cite{bauer18, silva19_1, galan20, baykusheva20, chacon20, schmid21, baykusheva21, silva18} has been explored. 

HHG research can be distinguished by material and laser wavelength: semiconductors exposed to (i) far-ir (far-infrared) \cite{schubert14,langer16,lanin17,huttner17} and (ii) mid-ir 
\cite{ghimire11,vampa15x,yoshikawa19,uzan20,wu17,dimitrovski17,dejean17,silva19_2} lasers, and (iii) dielectrics driven by near-ir lasers \cite{luu15,garg16,lakhotia20}. 

There is still considerable uncertainty about the various dominant mechanisms in the three experimental settings, impeding further development of HHG technology. The prevalent assumption 
is that HHG is driven by resonant mechanisms. Laser induced electron-hole pair creation is followed by HHG via interband polarization buildup in (ii) 
\cite{vampa15x,uzan20,vampa14,vampa15t,li19,yue20,parks20}, or via the intraband nonlinearity of individual bands in (iii) \cite{luu15,garg16}.

The contribution of virtual (non-resonant) nonlinear processes to HHG in solids, and to extreme nonlinear optics (NLO) in general, has been disregarded so far, despite their ubiquitous
importance in perturbative NLO \cite{boyd03}; virtual population results from laser induced distortion of the initial state and returns to the ground state after the laser pulse. Recent 
mid-ir semiconductor HHG experiments have indicated the importance of virtual processes for below minimum bandgap harmonics \cite{sanari20}. As such, the role of resonant versus virtual 
processes needs to be clarified for a more complete understanding of HHG. While resonant and virtual processes can be separated in perturbative NLO, this has not been possible in extreme 
NLO to date and presents uncharted territory. 

Here, closed form expressions for resonant and virtual nonlinear processes are obtained that are valid from the perturbative to the extreme realms of NLO, thus allowing a unified 
description of resonant and virtual NLO. 

Our approach generalizes the adiabatic following approach \cite{boyd03} from perturbative NLO in atoms to extreme NLO in solids. It results in closed-form equations for resonant and 
virtual interband and intraband currents driving HHG in solids. The validity of the strong field adiabatic following (SFAF) method is tested using 1D, two-band models for semiconductors 
and dielectrics. Excellent agreement between the exact von Neumann and the SFAF equations is found. 

The SFAF approach is applied to analyzing (ii) mid-ir semiconductor and (iii) near-ir dielectric HHG. The results for (ii) agree with and expand on previous findings. Resonant/non-resonant
interband HHG dominate above/below bandgap, and intraband HHG is negligible. HHG in the near-ir dielectric model system (iii) is dominated by a mixture of resonant and non-resonant interband 
and non-resonant intraband currents. The increasing importance of virtual mechanisms arises from a more pronounced role of the dynamic Stark shift in (iii), weakening resonant transitions. 
Beyond that, comparison of our results to experiments in (iii) reveals limitations of the so far primarily used microscopic one-electron-hole models. This charts the way towards a more 
complete description of HHG in solids. 

Our analysis starts from the von Neumann (one-body semiconductor Bloch) equation for density matrix $\rho$, 
\begin{align}
i \partial_t \rho(\mathbf{K},t) = [H(\mathbf{K}_t,t) , \rho(\mathbf{K},t)] \text{,} 
\label{vN}
\end{align}
derived in the moving crystal momentum frame $\mathbf{K}_t = \mathbf{K}+\mathbf{A}(t)$, with crystal momentum $\mathbf{K}$ defined in the shifted first Brillouin zone, 
$\overline{\text{BZ}} = \text{BZ} - \mathbf{A}(t)$. The vector potential is $\mathbf{A}(t)$ and the electric field $\mathbf{F}(t) = - \partial_t \mathbf{A}(t)$.  
Electron-hole interaction is neglected. We use atomic units, unless specified otherwise. The Hamilton operator $H$ is given by \cite{mcdonald15}
\renewcommand\arraystretch{1.3}
\begin{align}
H(\mathbf{K}_t,t) = \frac{1}{2} \begin{bmatrix}
\varepsilon(\mathbf{K}_t) & \Omega^*(\mathbf{K}_t,t) \\
\Omega(\mathbf{K}_t,t) & -\varepsilon(\mathbf{K}_t) 
\end{bmatrix} \text{}
\label{H}
\end{align}
with $\Omega(\mathbf{K}_t,t) = 2 \mathbf{F}(t) \mathbf{d}_{vc}(\mathbf{K}_t)$ the Rabi frequency and $ \mathbf{d}_{vc}$ the transition dipole element between valence 
and conduction band. We confine our analysis to inversion symmetric materials with real $ \mathbf{d}_{vc}$. The bandgap $\varepsilon(\mathbf{K}) = E_c(\mathbf{K}) - E_v(\mathbf{K})$ 
is the difference between conduction and valence energy bands $E_c, E_v$, respectively. Density operator and Hamiltonian are defined with respect to the Bloch basis functions $ \vert v 
\rangle(\mathbf{K}_t)$, $\vert c \rangle(\mathbf{K}_t)$, for valence and conduction band, respectively; e.g. $H_{11}$ refers to basis $\vert c \rangle \langle c \vert$. 

In the limit of laser frequency much smaller than the minimum bandgap, the electron dynamics follows dominantly the laser field and the adiabatic following approximation can be used. 
This is done by first diagonalizing the Hamiltonian (\ref{H}), 
\begin{align}
\widetilde{H}(\mathbf{K}_t,t) = V^+ H V = \frac{1}{2} \begin{bmatrix}
\lambda(\mathbf{K}_t,t) & 0 \\
0 & -\lambda(\mathbf{K}_t,t) 
\end{bmatrix} \text{,}
\label{Hdiag}
\end{align}
with $\lambda(\mathbf{K}_t,t) = \sqrt{\varepsilon^2(\mathbf{K}_t) + \Omega^2(\mathbf{K}_t,t)}$ and unitary matrix 
\begin{align}
V(\mathbf{K}_t,t) = \frac{1}{\sqrt{2}}\begin{bmatrix}
\frac{\sqrt{\lambda+\varepsilon}}{\sqrt{\lambda}} & -\frac{\Omega}{\sqrt{\lambda} \sqrt{\lambda+\varepsilon}} \\
\frac{\Omega}{\sqrt{\lambda} \sqrt{\lambda+\varepsilon}} & \frac{\sqrt{\lambda+\varepsilon}}{\sqrt{\lambda}} 
\end{bmatrix} \text{.}
\label{V}
\end{align}

Multiplying the von Neumann equation with $V^+, V$ from the left and right, inserting $V V^+ = \mathbb{1}$ and defining $\tilde{\rho} = V^+ \rho V$ yields
\begin{align}
i \partial_t \tilde{\rho} = [\widetilde{H}, \tilde{\rho}] + i ({\partial_t V^+}) V \tilde{\rho} + i \tilde{\rho} V^+ ({\partial_t V}) \text{.}
\label{vN1}
\end{align}
Integration of Eq. (\ref{vN1}) and transformation back to $\rho$ yields 
\begin{align}
& \rho(\mathbf{K},t) = X^+(\mathbf{K},t) \rho(\mathbf{K},t=-\infty) X(\mathbf{K},t) \text{} \label{rhox} \\  
& X(\mathbf{K},t) = \left( \hat{T} e^{ \int_{\!_{-\infty}}^{t} \!\!\!\!\! d\tau W(\mathbf{K},{\tau}) } \right)^{\!+} \!\! e^{i \int_{\!_{-\infty}}^{t} \!\!\!\!\! d\tau 
\widetilde{H}(\mathbf{K}_{\tau})} V^+(\mathbf{K}_t) \text{} \nonumber \\
& W(\mathbf{K},t) \! = \! e^{i \int_{\!_{-\infty}}^{t} \!\!\!\!\! d\tau \widetilde{H}(\mathbf{K}_\tau) } (\partial_t V^+(\mathbf{K}_t)) V(\mathbf{K}_t) 
e^{- i \int_{\!_{-\infty}}^{t} \!\!\!\!\! d\tau \widetilde{H}(\mathbf{K}_\tau) }
\nonumber
\end{align}
For the sake of brevity, we omit 
the explicit time dependence in $V, V^+, \widetilde{H}, \Omega$, and $\lambda$ from Eq. (\ref{rhox}) onward. 
Here, $X(\mathbf{K},t)$ and $W(\mathbf{K},t)$ are matrix operators, and $\hat{T}$ refers to the time ordering operator, which numerically is evaluated as $\hat{T} 
e^{ \int_{\!_{-\infty}}^{t} \!\!\!\!\!\! d\tau W(\mathbf{K},\tau)} = \Pi_{j=0}^{n} e^{W(\mathbf{K},t_j) d\tau}$ on a time window between $t_0$ and $t_n$ with step size 
$d\tau \rightarrow 0$ small enough to converge. The time ordered operator can be expanded into a Dyson series \cite{sakurai11}; keeping terms up to second order yields
\begin{align}
& \rho(\mathbf{K},t) = \rho_0(\mathbf{K},t) + \rho_1(\mathbf{K},t) + \rho_2(\mathbf{K},t) \approx \label{rho} \\
& \approx \begin{bmatrix}
0 & 0 \\
0 & 1 
\end{bmatrix} + 
\begin{bmatrix}
0 & - \frak{u} \\
- \frak{u}^* & 0  
\end{bmatrix}_{\mathbf{K},t}  \!\!\!\!\!\!\! + 
\begin{bmatrix}
\vert \frak{u} \vert^2 & 0 \\
0 & - \vert \frak{u} \vert^2  
\end{bmatrix}_{\mathbf{K},t} \,\,\, \textrm{with} \nonumber \\
& \frak{u}(\mathbf{K},t) = \frac{i}{2} \int_{-\infty}^t \!\!\!\!\!\! dt' \Omega(\mathbf{K}_{t'}) e^{-i\int_{t'}^t \lambda(\mathbf{K}_{\tau}) d\tau } \text{.}
\label{ux}
\end{align}
Here, we have assumed $\Omega / \varepsilon \le 1$ and neglected terms of $O(\Omega / \varepsilon)$ and higher in the preexponent. A detailed derivation will be given elsewhere 
\cite{neda}.

HHG is determined by the absolute square of the Fourier transform ($\text{FT}$) of the expectation value of the current, $ \vert \text{FT} \{ \langle \mathbf{j} \rangle (t) \} \vert^2$; 
the current expectation value is given by $\langle \mathbf{j} \rangle(\mathbf{K},t) = \int_{\overline{\text{BZ}}} d^3\!K \, \text{Tr}[\rho(\mathbf{K},t) \mathbf{j}(\mathbf{K}_t)]$. 

The current operator is represented by $\mathbf{j}_{jl} = -i \langle j \vert \boldsymbol{\nabla} \vert l \rangle$ with $\vert j \rangle, \vert l \rangle = \vert 
c\rangle(\mathbf{K}_t),\vert v \rangle(\mathbf{K}_t)$. 
Here, the diagonal elements $\mathbf{j}_{jj} = \boldsymbol{\nabla_{\mathbf{K}}} E_j = \mathbf{v}_j(\mathbf{K})$ represent the band velocities and the bandgap velocity is defined as 
$\mathbf{v}(\mathbf{K}) = \mathbf{v}_c(\mathbf{K}) - \mathbf{v}_v(\mathbf{K})$. The off-diagonal terms $ \mathbf{j}_{jl} \rho_{lj} = (d/dt) \mathbf{d}_{jl}(\mathbf{K}_t) 
\rho_{lj}(\mathbf{K},t)$ are expressed in terms of the interband dipole moment $\mathbf{d}_{jl}$. 

The current expectation value can be decomposed into contributions coming from the various density matrix expansion orders, $\langle \mathbf{j} \rangle = \sum_{j=0}^2 \langle \mathbf{j}_j 
\rangle$, where $\langle \mathbf{j}_j \rangle = \int_{\overline{\text{BZ}}} d^3\!K \, \text{Tr}[\rho_j(\mathbf{K},t) \mathbf{j}(\mathbf{K}_t)]$. We only get HHG contributions from 
$ \langle \mathbf{j}_{1} \rangle$ and $ \langle \mathbf{j}_{2} \rangle$. Replacing $\langle \mathbf{j}_{1} \rangle \rightarrow \langle \mathbf{j}_{er} \rangle$ and $\langle \mathbf{j}_{2} 
\rangle \rightarrow \langle \mathbf{j}_{ra} \rangle$ we obtain interband and intraband current 
\begin{subequations}
\label{TrJs}
\begin{align}
\langle \mathbf{j}_{er} \rangle & \approx - \frac{d}{dt} \int_{\overline{\text{BZ}}} d^3\!K \, \mathbf{d}_{vc}(\mathbf{K}_{t}) \, \frak{u}(\mathbf{K},t) 
+ \text{c.c.} \text{,} 
\label{TrJs1} \\
\langle \mathbf{j}_{ra} \rangle & \approx \int_{\overline{\text{BZ}}} \! d^3\!K  \mathbf{v}(\mathbf{K}_{t}) n_c(\mathbf{K},t) \text{.}
\label{TrJs2} 
\end{align}
\end{subequations}
Here, $n_c(\mathbf{K},t) = \vert \frak{u}(\mathbf{K},t) \vert^2$ and $n_c(t) = \int_{\overline{\text{BZ}}} d^3\!K n_c(\mathbf{K},t)$ is the conduction band population. 
In the limit of small Rabi frequency, $\lambda \rightarrow \varepsilon$ in the exponent of Eq. (\ref{ux}), Eqs. (\ref{TrJs}) go over into the frozen valence band (FVB) solution 
\cite{vampa14,mcdonald17}. Here we observe one of the key findings revealed by this approach; the main difference between the SFAF and FVB solution is the dynamic Stark shift. 

\begin{figure}[t]
\includegraphics[width=8.3cm]{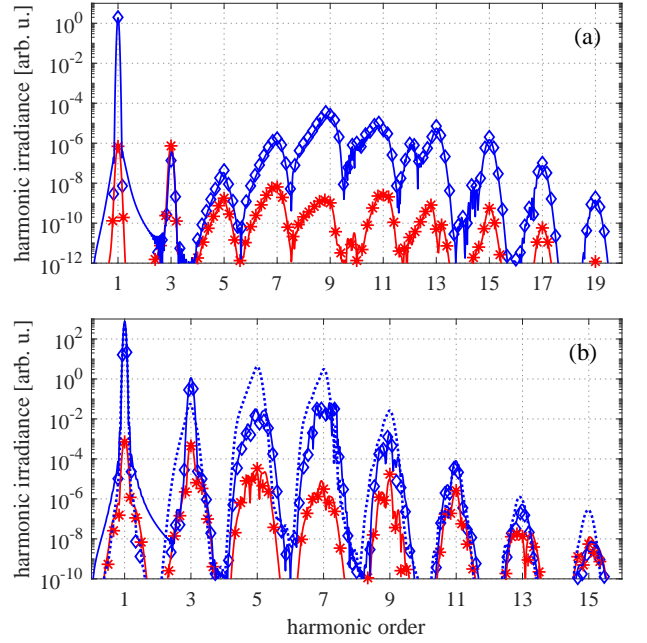} 
\caption{\label{fig1} (a,b) Interband (blue full line) and intraband (full red line) HHG as obtained from numerical solution of Eq. (\ref{vN}) are compared to interband (blue diamond) 
and intraband (red asterisk) HHG from SFAF Eqs. (\ref{TrJs}). Dotted line in (b) represents FVB solution ($\lambda \rightarrow \varepsilon$) in Eq. (\ref{ux}). (a) Model semiconductor: 
$E_g = 0.129$, $\Delta_1 = 0.17$, $d_0 = 3.64$; mid-ir laser: $F_0 = 0.002$ ($1.2 \times 10^{11}$W/cm$^2$), $\omega_0 = 0.015$ ($3.04\mu$m), $\tau_0 = 6T_0$. (b) Model dielectric: 
$E_g = 0.32$, $\Delta_1 = 0.06$, $\Delta_{2} = -0.0035$, $\Delta_{3} = -0.001$, $\Delta_{4} = -0.0007$, $d_0 = 6.5$; near-ir laser: $F_0 = 0.012$ ($4.3 \times 10^{12}$W/cm$^2$), 
$\omega_0 = 0.06$ ($0.76\mu$m), 
$\tau_0 = 6T_0$.}
\end{figure}

The validity of Eqs.\,(\ref{TrJs}) is verified by using a two-band, 1D model solid driven by a laser pulse $F(t) = F_0 \exp(-(t/\tau_0)^2) \sin(\omega_0 t)$. Here, $F_0$ is the 
peak electric field strength, $\omega_0$ is the laser frequency (related the the laser period $T_0$ as $\omega_0=2\pi/T_0$), and $\tau_0$ is the Gaussian halfwidth. The bandgap is given by $\varepsilon(K) = E_g + \sum_{j=1}^n \Delta_n (1-\cos(nKa))$,
where $E_g$ is the minimum bandgap and the halfwidth of the bandgap is mainly determined by $\Delta_1$. The dipole moment in $\mathbf{K} \cdot \mathbf{p}$ approximation is given by $d_{vc} 
= d_0\varepsilon(K=0) / \varepsilon(K)$. We investigate two model systems representative of mid-ir semiconductor and near-ir dielectric experiments with $\Delta_1/\omega_0 \gg 1, \sim 1$,
respectively. For parameters see Fig.\,\ref{fig1}. 

In Fig.\,\ref{fig1} interband HHG (blue line) and intraband HHG (red line), as determined by a numerical solution of Eq.\,(\ref{vN}) for model semiconductor (a) and dielectric (b), are 
compared to harmonic spectra obtained from currents (\ref{TrJs1}) (blue diamonds) and (\ref{TrJs2}) (red asterisks). 
The blue dotted line in (b) represents FVB interband HHG. It is not shown in (a), as it overlaps with the exact numerical solution. Interband HHG is dominant over the whole spectrum in 
(a). By contrast, there is a difference of up to two orders between the SFAF and FVB results for interband HHG; FVB intraband HHG is not shown, but displays similar disagreement. The 
greater importance of the dynamic Stark effect in dielectrics can be attributed to larger dipole moments and to higher applicable intensities due to higher damage thresholds. Finally, 
for higher harmonics, $N \ge 15$, the spectral intensities of inter- and intraband contributions become comparable in Fig.\,\ref{fig1}(b); see also Fig.\,\ref{fig4}. 

In the remainder, resonant and non-resonant contributions to HHG are isolated from the SFAF Eqs. (\ref{TrJs}) and analyzed. This is done by first splitting Eq.\,(\ref{ux}) into a
probability amplitude of ionization, $\frak{v}$, and into an exponent that is responsible for interband HHG. Then, $\frak{v} = \frak{v}_{r} + \frak{v}_{nr}$ and $\frak{u} = \frak{u}_r + 
\frak{u}_{nr}$ are split into resonant and non-resonant parts,  
\begin{align}
\frak{u}_{i}(\mathbf{K},t) & = e^{-i\int_{\!_{-\infty}}^t \!\!\!\!\! d\tau \lambda(\mathbf{K}_{\tau}) } \frak{v}_{i}(\mathbf{K},t) \,\,\, (i=r,nr) \text{,} \label{frakui}
\end{align}
see schematic \ref{fig3}(a). Whereas resonant transitions are expected to exhibit a steady increase of $n_c$ over time, non-resonant transitions are oscillatory and all of the population 
returns to the valence band after the laser pulse. Mathematically, this translates into the resonant filter $G_r(\omega) = 1$ for $-\omega_0/2 \le \omega \le \omega_0/2$ and 
$G_r(\omega) = 0$ elsewhere; the nonresonant filter is $G_{nr}(\omega) = 1 - G_r(\omega)$. Therefrom, the resonant (non-sinusoidal) and nonresonant (sinusoidal) transition probability 
amplitudes are given by $\frak{v}_{i}(\mathbf{K},t) = \textrm{FT}^{-1} G_{i}(\omega) \tilde{\frak{v}}(\mathbf{K},\omega)$ for $i=r,nr$, respectively. To separate the intraband current, 
$n_c = \vert \frak{v} \vert^2 = n_c^r + n_c^{nr}$ also needs to be split with $n_c^i(t) = \int_{\overline{\text{BZ}}} d^3\!K n_c^i(\mathbf{K},t)$ ($i = r,nr$),  
\begin{align}
n_c^r(\mathbf{K},t) & = \vert \frak{v}_r(\mathbf{K},t) \vert^2 \label{ncr} \\
n_c^{nr}(\mathbf{K},t) & = \vert \frak{v}_{nr}(\mathbf{K},t) \vert^2 + \left[ \frak{v}_{nr}(\mathbf{K},t) \frak{v}_{r}^*(\mathbf{K},t) + \text{c.c.} \right] = \nonumber \\ 
                       & = n_c^{n_1}(\mathbf{K},t) + n_c^{n_2}(\mathbf{K},t) \text{.} \label{ncnr} 
\end{align}
The squared bracket in Eq.\,(\ref{ncnr}) corresponds to $n_c^{n_2}$. 

\begin{figure}[t]
\includegraphics[width=8.6cm]{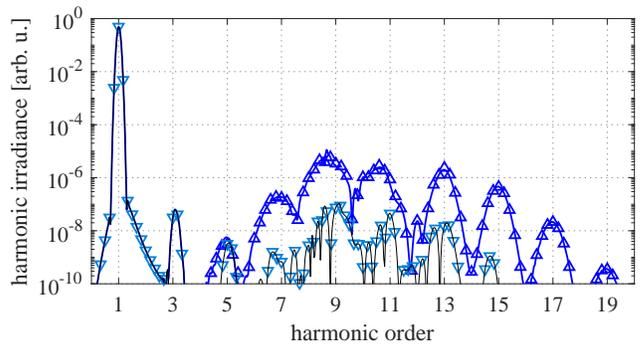} 
\caption{\label{fig2} Model semiconductor with same parameters as in Fig.\,\ref{fig1}(a); HHG from $j_{er}$ (full blue line), $j_{er}^{r}$ (blue triangles up), and $j_{er}^{nr}$ 
(blue triangles down, different shade of blue for visibility; thin black line is a guide to the eye) is compared. Symbols are plotted with lower resolution.}
\vspace*{-0.3cm}
\end{figure}

With the above definitions, the resonant and non-resonant interband and intraband currents are 
\begin{align}
\langle \mathbf{j}_{er}^{i} \rangle & \approx - \frac{d}{dt} \int_{\overline{\text{BZ}}} d^3\!K \, \mathbf{d}_{vc}(\mathbf{K}_{t}) \, \frak{u}_i(\mathbf{K},t) 
+ \text{c.c.} \text{,} \label{jeri} \\
\langle \mathbf{j}_{ra}^{i} \rangle & \approx \int_{\overline{\text{BZ}}} \! d^3\!K  \mathbf{v}(\mathbf{K}_{t}) n_c^{i}(\mathbf{K},t) \,\,\, (i = r,nr) \text{,} \label{jrai} 
\end{align}
where, according to Eq.\,(\ref{ncnr}), the non-resonant intraband current consists of two parts $\langle \mathbf{j}_{ra}^{nr} \rangle = \langle \mathbf{j}_{ra}^{n_1} \rangle + 
\langle \mathbf{j}_{ra}^{n_2} \rangle$. 

Now, Eqs.\,(\ref{frakui})\,-\,(\ref{jrai}) are applied to the model systems of Fig.\,\ref{fig1} to separate resonant from non-resonant HHG. The markers in 
Figs.\,\ref{fig2}\,-\,\ref{fig4} reflect the relation between HHG currents, see Ref. \cite{comment}. For $n_c^{i}$ the same symbols are used as for $j_{ra}^{i}$ ($i=r,nr$). 

In Fig.\,\ref{fig2} the interband harmonic spectrum from Fig.\,\ref{fig1}(a) is split into resonant and non-resonant contributions. HHG from the full interband current\,(\ref{TrJs1}) 
(blue full line), and from resonant (blue triangles up) and non-resonant interband currents (blue triangles down) in Eq.\,(\ref{jeri}) is compared; HHG for $N=1,3$ is non-resonant 
and turns resonant for $N>5$; the first above bandgap harmonic is $N=9$. Above minimum bandgap harmonics are resonant, in agreement with previous theoretical 
\cite{vampa14, vampa15t, li19, yue20, parks20} and experimental \cite{vampa15x, uzan20, vampa20} work. 

\begin{figure}[t]
\includegraphics[width=8.6cm]{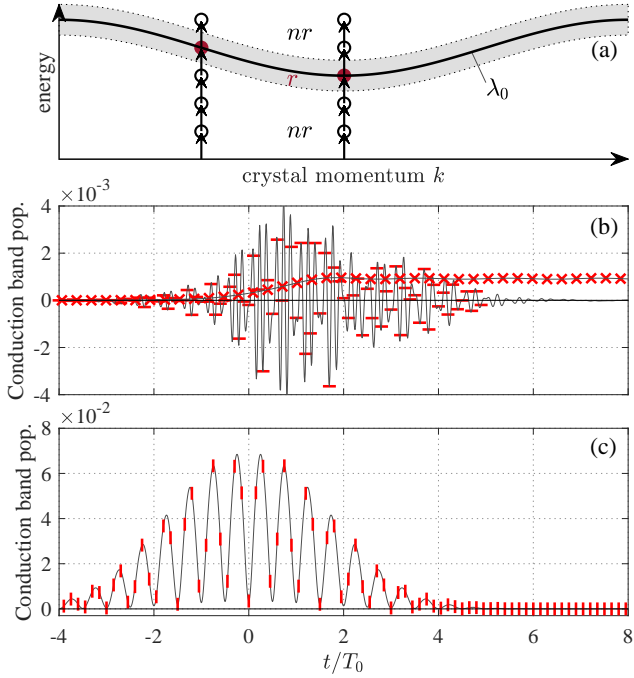} 
\caption{\label{fig3} (a) schematic of ionization amplitude $\frak{v}$. In the continuous wave (cw)-limit, a Fourier series expansion of the integrand of $\frak{v}$ yields 
$\lim_{t\rightarrow \infty} \frak{v} = \sum_n \int_{0}^{\infty} dt' c_n e^{i (\lambda_0(\mathbf{K}) + n \omega_0) t'} = \sum_n  c_n(\mathbf{K}) \left\{ i \pi \delta(\lambda_0(\mathbf{K}) 
+ n \omega_0) + \textrm{P} (1/(\lambda_0(\mathbf{K}) + n \omega_0)) \right\} = \frak{v}_r + \frak{v}_{nr}$ with the non-sinusoidal term $\lambda_0(\mathbf{K})$ a field dressed bandwidth 
(black line), and $c_n$ Fourier coefficients. The $\delta$ function and principal value integral $\textrm{P}$ represent resonant ($r$) (purple, full circle) and non-resonant ($nr$)
(empty, black circles) transitions for each $K$-value, respectively. In going from cw to finite pulses, the black line morphs into the grey shaded area of width $\pm \omega_0/2$. For the
resulting splitting procedure see text. 
(b,c) Conduction population time evolution for the model dielectric; parameters are the same as in Fig.\,\ref{fig1}(b) except for $\tau_0 = 3 T_0$ and $F_0 = 0.02$. (b) $n_c^{r}$ 
(Eq.\,(\ref{ncr}), red cross) and $n_c^{n_2}$ (Eq.\,(\ref{ncnr}), red horizontal line); (c) $n_c^{n_1}$ (Eq.\,(\ref{ncnr}), red vertical line); thin black lines: zero lines and guide to 
the eye.}
\vspace*{-0.35cm}
\end{figure}

In the model dielectric of Fig.\,\ref{fig1}(b) intraband currents $j_{ra}^{i}$ and thereby populations $n_c^i$ ($i = r,n1,n2$) become relevant; $n_c^r$ (red crosses) and $n_c^{n2}$ 
(red horizontal bars) are plotted in Fig.\,\ref{fig3}(b); for $n_c^{n1}$ (red vertical lines) see Fig.\,\ref{fig3}(c). The parameters are the same as in Fig.\,\ref{fig1}(b) except for 
$F_0 = 0.02$ and a shorter pulse duration $\tau = 3 T_0$ usually used in high intensity experiments. As required, the resonant contribution keeps steadily growing and reaches a constant 
value after the laser pulse. Usually, tunnel ionization calculations display a sub-cycle, step-like increase arising from the exponential sub-cycle field dependence. Our formalism counts 
the step-like increase as part of the non-resonant oscillations about the (cycle-averaged) resonant ionization curve displayed in Fig.\,\ref{fig3}(b). The non-resonant, virtual 
contribution $n_c^{n_1}$ is oscillatory and goes to zero after the pulse. Finally, the mixed, resonant-nonresonant contribution $n_c^{n_2}$ also goes to zero after the pulse, revealing 
its virtual nature. It is smaller but more rapidly oscillating than $n_c^{n_1}$ which is why it can also substantially contribute to HHG. 

\begin{figure}[t]
\includegraphics[width=8.6cm]{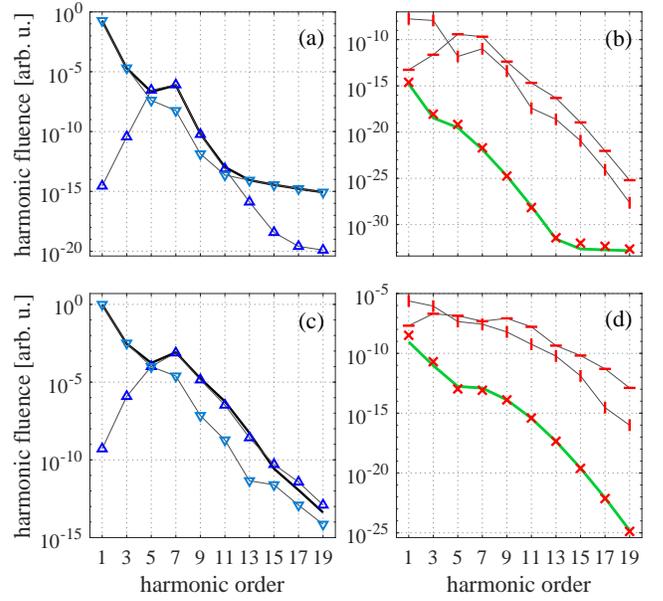} 
\caption{\label{fig4} same parameters as in Fig.\,\ref{fig3} except for $F_0 = 0.005$ (a,b) and $F_0 = 0.012$ (c,d). Thin gray lines serve as guide to the eye. (a,c) HHG from interband 
currents $j_{er}^{r}$ (blue triangles up), and $j_{er}^{nr}$ (blue triangles down), HHG from total current (black full line). (b,d) HHG from intraband currents $j_{ra}^{r}$ (red crosses), 
$j_{ra}^{n_1}$ (red vertical lines), $j_{ra}^{n_2}$ (red horizontal lines), $j_{ra}^{x}$, see text (green full lines).}
\vspace*{-0.35cm}
\end{figure}

HHG in the model dielectric is plotted in Fig.\,\ref{fig4} for $F_0 = 0.005$ (a,b) and for $F_0 = 0.012$ (c,d); the remaining parameters are as in Fig.\,\ref{fig3}. Sub-plots (a,c) and 
(b,d) show interband and intraband currents, respectively. For clarity, we plot the harmonic fluence by integrating harmonic signals over the frequency interval $|\omega-N\omega_0| < 
(\omega_0/2)$. The first above minimum bandgap harmonic is $N=7$. In (a,b) the total harmonic signal (black full line) is composed of a mixture of $j_{er}^{r}$ (blue triangles up) 
and $j_{er}^{nr}$ (blue triangles down); intraband currents (b) are negligible; the resonant intraband current (red crosses) is weakest. For higher $F_0$, Figs.\,\ref{fig4}(c,d), $j_{er}^{r}$ 
(blue triangles up) is dominant for $N>5$ and accounts for most of the total harmonic signal (full black line) in (c). For $N \ge 11$ $j_{ra}^{n_2}$ (red horizontal lines) becomes 
comparable to $j_{er}^{r}$. Again, $j_{ra}^{r}$ (red crosses) is weakest. This explains why HHG from $j_{ra}^{r}$ could not be observed in numerical analysis of the 
semiconductor Bloch equations \cite{luu15, garg16}. Finally, this presents the first direct identification of the relevance of virtual HHG channels. Their greater importance in near-ir
dielectric experiments can be attributed to the dynamic Stark effect, which increases the effective minimum bandgap and thus weakens resonant transitions \cite{mcdonald17}.

Beyond that, another key finding ensues from comparison to experiments; it reveals first evidence that processes going beyond the microscopic one-electron-hole approach are important.
Near-ir dielectric HHG \cite{luu15, garg18, lakhotia20} can be explained quite well by a simple model (Bloch oscillation) current $j_{ra}^{x} = c \vert \textrm{FT}[v(A(t))] 
\vert^2$, where $c$ is a constant. It can be obtained from Eq.\,(\ref{jrai}) by assuming that the conduction band population is delta-function like around $K=0$. It is represented by the 
green line in Figs.\,\ref{fig4}(b,d), which has been matched to $j_{ra}^{r}$ by determining $c$ at harmonic $N=9$. The constant is found to be of the order of $n_c^2(t\rightarrow \infty)$. 
The agreement between $j_{ra}^{r}$ and $j_{ra}^{x}$ is excellent in both cases. Note, that the (cycle-averaged) resonant ionization rate does not contain harmonic terms so that resonant
intraband HHG comes solely from the band velocity. 

The negligible contribution of $j_{ra}^{r,x}$ to the total current in Fig.\,\ref{fig4} conflicts with the fact that $j_{ra}^{x}$ was successfully used to describe near-ir experiments 
in dielectrics \cite{luu15} and below bandgap harmonics in some semiconductors \cite{lanin17}. The apparent contradiction indicates that important physics is missing in semiconductor 
Bloch analysis. An important missing element is dephasing, be it through propagation \cite{floss18, kilen20, jurgens20}, or through microscopic scattering processes. Dephasing can strongly
affect the oscillatory parts of $e^{i \int_{t'}^t \lambda d\tau}$ in Eq.\,(\ref{ux}), thus, suppressing HHG. The presence of dephasing should favor $j_{ra}^{r}$, as it is the only 
non-oscillatory current that changes on the time-scale of the laser envelope. Our preliminary 3D studies lead to the same qualitative conclusions. Inclusion of higher bands is subject to 
future research.

Finally, the method developed here lays the necessary theoretical foundation to pursue the above ideas. The SFAF equations can be integrated very efficiently and are ideally suited for 
coupling with heat baths and with macroscopic propagation equations.

\section*{Author contributions}
A. Thorpe and N. Boroumand are joint first authors for this work.

\end{document}